\documentclass{aa}

\usepackage{graphicx}
\usepackage{txfonts}
\usepackage{braket}
\usepackage{amsmath}
\usepackage{bm}
\usepackage[colorlinks,allcolors=blue]{hyperref}
\usepackage[a]{esvect}
\makeatletter
\renewcommand*\aa@pageof{, page \thepage{} of \pageref*{LastPage}} 

\makeatother
\usepackage{amstext}
\usepackage[normalem]{ulem}

\begin{document}

\title{Adaptation of the phase distance correlation periodogram to account for measurement uncertainties}


\author{A. Binnenfeld\inst{\ref{inst1}} \and S. Shahaf\inst{\ref{inst2}} 
\and S. Zucker\inst{\ref{inst1}}}

\institute{Porter School of the Environment and Earth Sciences, Raymond and Beverly Sackler Faculty of Exact Sciences, 
Tel Aviv University, Tel Aviv, 6997801, Israel \\
\email{avrahambinn@gmail.com}\label{inst1}
\and
Department of Particle Physics and Astrophysics, Weizmann Institute of Science, Rehovot 7610001, Israel
\label{inst2}}

\date{Accepted XXX. Received YYY}

\abstract{
We present an improvement of the phase distance correlation (PDC) periodogram to account for uncertainties in the time-series data. The PDC periodogram introduced in our previous papers is based on the statistical concept of distance correlation. By viewing each measurement and its accompanying error estimate as a probability distribution, we are able to use the concept of energy distance to design a distance function (metric) between measurement-uncertainty pairs. We used this metric as the basis for the PDC periodogram, instead of the simple absolute difference. We demonstrate the periodogram's performance using both simulated and real-life data. This adaptation makes the PDC periodogram much more useful, demonstrating it can be helpful in the exploration of large time-resolved astronomical databases, ranging from Gaia radial velocity and photometry data releases to those of smaller surveys, such as APOGEE and LAMOST. We have made a public GitHub repository available, with a Python implementation of the new tools available to the community.
}


\titlerunning{PDC periodogram
accounting for measurement uncertainties}
\authorrunning{A. Binnenfeld et al.}

\maketitle
\section{Introduction}
\label{sec:intro}

Time-domain astronomy poses distinctive challenges, including sparse and uneven sampling, heteroscedasticity, high-dimensionality, and large data volumes. In practice, the astrophysical information obtained from an astronomical observation is often reduced to some scalar quantity, such as the radial velocity or photometric magnitude. The temporal behavior of these observables helps shed light on the system's physical properties. 

One key aspect of astronomical time series analysis is the search for periodic modulation in the observed data. A common approach to detecting periodicity is to generate a periodogram, produced by scanning a grid of trial periods (or frequencies) and ascribing each one with a score. The score quantifies the plausibility that the data are indeed modulated with a given period. A more thorough characterization of the observed object is usually initiated only if a periodic modulation is identified. The appeal of this approach stems from its conceptual simplicity and reduced computational costs.

The \textit{Lomb-Scargle} (LS) periodogram is a widely used, weighted, least squares-based method for periodicity detection in unevenly sampled data \citep{1981AJ.....86..619F}. It assumes that a sine wave modulation is able to explain most of the variation (`energy') in the data and that the measurement error is additive, uncorrelated ('white'), and Gaussian. If the underlying signal is sinusoid and the noise is white and Gaussian, LS would be an optimal detection scheme, based on the Neyman-Pearson hypothesis-testing theory. The logic behind its application largely relies on the ability to approximate smooth periodic functions by Fourier series, assuming that most energy will concentrate on one fundamental mode \citep[see][]{2018ApJS..236...16V}. 

However, the assumption that the periodic signal is well represented by a single harmonic function can be invalid in certain cases. This holds particularly true for light curves exhibiting the majority of photometric variability patterns. As a consequence, several extensions to LS have been developed, aiming to broaden the flexibility of the search. For example, \citet[][]{FAWCETT2006861} used several harmonics, instead of the pure sinewave used in the classical LS periodogram; another extension is attributed to \citet{GLSpaper}, where a constant offset term was incorporated to account for non-zero baseline values. The latter approach is often referred to as the generalized LS periodogram (GLS).  

Nevertheless, some astrophysical phenomena may give rise to signals that cannot be approximated with only a few harmonics. Some studies have shown that GLS and other parametric periodograms tend to underperform in such cases \citep{Pinamonti17, zucker18}. Examples of such signals include the radial-velocity curves of highly eccentric Keplerian orbits or sawtooth-like photometric modulations of pulsating stars. Exoplanetary transits are another well-known example of a type of signal that is poorly approximated by sinusoids. Instead of increasing the number of harmonics in the template model used for the search, the common approach is changing its functional shape altogether. The classical transit detection method BLS (box-fitting least Squares) looks for box-shaped signals in the data \citep{bls, panahi21, shahaf22}. As the photometric accuracy improved, the assumed underlying model was further refined in various ways to employ more realistic transit-shaped models \citep[e.g.][]{man_ag2002, Kipping2023}, further improving the search sensitivity. 

The approaches discussed above are parametric, as they rely on the ability to model or approximate the functional shape of the signal. There are cases, however, where the functional shape is unknown or requires significant computational resources for the calculation. Several non-parametric periodograms were developed to address such situations. These techniques are designed to be more agnostic to the particular modulation shape, expanding the reach of the search, sometimes at the cost of reduced statistical efficiency:

String-length techniques are a set of commonly used model-independent methods \citep{clarke2002}. They include, among others, the Lafler-Kinman method \citep{laki1965}, as well as the methods of \citet{ren1978} and \citet{dwo1983}. In all of those methods, we quantify the dependence between consecutive phase-folded measurements by estimating the length of an imaginary string connecting them. 

The analysis of variance method \citep[AoV;][]{aov89} is another non-parametric approach, which basically fits a periodic piecewise constant function to the measurements. After dividing the measurements into phase subsets ('bins') based on the trial period, the variance within each subset is compared to the variance of the entire dataset to determine the significance of the periodic signal at each period. A method closely related to the AoV method is the phase dispersion minimization method \citep[PDM;][]{Stellingwerf78}, minimizing the dispersion of the phase-folded data sets to detect the most probable period.

The phase distance correlation periodogram (PDC; \citealt{zucker18}) quantifies the statistical dependence between the observable and the phase at which the data were taken, given the trial period. The statistical dependence is quantified by the distance correlation \citep{Szeetal2007}. The PDC periodogram does not assume a specific shape for periodic variability and outperforms GLS in detecting some highly non-sinusoidal periodic signals \citep{zucker18}. 

One weakness of the non-parametric methods discussed above is their inability to consider measurement uncertainties. Error estimates are essential components of all physical measurements \citep[e.g.][]{barlow1995statistics}. Uncertainties can reflect various noise sources, such as inherently unavoidable photon noise or instrumental glitches, and their estimates are particularly fundamental for understanding and analyzing astronomical measurements \citep[see][]{Andrae2010}.

Uncertainties constitute a unique challenge for model-independent periodograms such as the PDC. Without an assumed functional shape, it is not straightforward to define, let alone quantify, the deviations of the data from their expected value. As a result, uncertainties are often omitted from the calculation in these cases. One notable exception is related to the AoV and PDM methods mentioned earlier, where an assumed functional form does indeed exist, specifically one that minimizes dispersion. In this case, the minimization procedure can involve sampling the assumed distribution of the noise to incorporate them into the procedure.

In this work, we present an improvement of the PDC periodogram that considers measurement uncertainties. We do so by viewing each measurement paired with its corresponding uncertainty as a probability distribution, and then introducing a new metric that enables the calculation of distance correlation and therefore the PDC periodogram. 

In the next section, we present the details of adapting the PDC periodogram to account for measurement uncertainties. In Sect.~\ref{sec:ex2}, we demonstrate the performance of the improved periodogram using simulated data and its application to the HARPS data of a planet-hosting star. We present our conclusions in Sect.~\ref{sec:conc} and discuss the method and its potential for future studies.

\section{Incorporating measurement errors in PDC}
\label{sec:pdc}

\subsection{A new metric}
As mentioned above, the distance correlation is a way to quantify the statistical dependence between two random variables \citep{Szeetal2007}. One of its merits is that the two random variables in question do not need to be of the same dimensionality \citep{Lyons2013}. 

Distance correlation is reliant on providing each random variable with an adequate metric to determine the pairwise distance of the sample (\citealt{Lyons2013}). 
Therefore, to allow the PDC periodogram to account for measurement uncertainties, we must define a suitable metric to consider both the measurements and their accompanying error estimates. To do so, we assume that each measurement is sampled from a Gaussian distribution, with the nominal measurement value as its mean ($\mu$) and the accompanying error estimate as its standard deviation ($\sigma$). We define the distance between two measurements as the `Energy Distance' between their two corresponding distributions \citep{Szkely2013EnergySA}. This metric satisfies the required conditions, outlined by \citet{Rizzo2016}, for the distance correlation to be applied as a measurement of statistical independence.

The distance between two measurements is taken as the energy distance between the two probability distributions from which these values were sampled, $F_i$ and $F_j$. The explicit expression for the energy distance is \citep{Szkely2013EnergySA}:
\begin{equation}
\label{EnergyDistance}
\mathcal{D}^2(F_i, F_j) = 2E||X_i - X_j|| - E||X_i - X_i'|| - E||X_j - X_j'||\ ,
\end{equation}
where $X_i$ and $X_j$ are independent random variables distributed according to $F_i$ and $F_j$; $X_i'$ and $X_j'$ are independent and identically distributed copies of $X_i$ and $X_j$; $||\cdot||$ denotes the Euclidean norm; $E$ represents the expectancy-value. 

The difference between two normally distributed random variables is also normally distributed. Therefore, $X_i - X_j$ is a Gaussian random variable with mean $\mu_i - \mu_j$ and variance $\sigma_i^2 + \sigma_j^2$. Similarly, $X_i - X_i'$ and $X_j - X_j'$ are zero-mean Gaussian distributions with variances given by $2\sigma_i^2$ and $2\sigma_j^2$. Finally, to obtain an expression for the energy distance, we need to know the distribution of the random variable $||X_i - X_j||$. Since we are considering random scalars, the last expression is a random variable representing the absolute value of the difference between two normally distributed variables, which follows a `folded normal distribution', which is the distribution of the absolute value of a normally distributed random variable.

\citet{FoldedNormal} showed that the expectancy value of a folded normal distribution is
\begin{equation}
\sigma \sqrt{\frac{2}{\pi}} \exp\left({-\frac{\mu^2}{2\sigma^2}}\right) + \mu\, \operatorname{erf}\left(\frac{\mu}{\sqrt{2\sigma^2}}\right) \ ,
\end{equation}
where $\mu$ and $\sigma^2$ represent the mean and variance of the underlying Gaussian, whose absolute value was taken, and $\operatorname{erf}$ is the Gauss error function, as follows:
\begin{equation*}
\operatorname{erf}\left(x\right) = \frac{2}{\sqrt{\pi}} \int_0^x \exp\left(-t^2\right) dt \ .
\end{equation*}

With these results in hand, we can explicitly write the energy distance (Eq.~\ref{EnergyDistance}) for two normally distributed random variables. For convenience, we define $\sigma_{ij}^2\equiv {{\sigma_i^2+\sigma_j^2}}$. In these terms, Eq.~\ref{EnergyDistance} becomes 
\begin{equation}
\begin{aligned}
\varepsilon^2_{ij} \, = \,&  {\sqrt{\frac{8}{\pi}}} \,\sigma_{ij} \, \left[{\exp({-x^2})} +  x\,\operatorname{erf}\left(x\right) 
-y \right] \ ,
\end{aligned}
\label{dist_eq}
\end{equation}
where
\begin{equation*}
\begin{aligned}
    x\equiv \, \frac{\mu_i-\mu_j}{\sqrt{2}\sigma_{ij}} \quad \text{and} \quad
    y \equiv \, \frac{\sigma_i+\sigma_j}{\sqrt{2}\sigma_{ij}}\, .
\end{aligned}
\end{equation*}

\begin{figure} 
\centering
\includegraphics[width=0.85\columnwidth,clip=true]{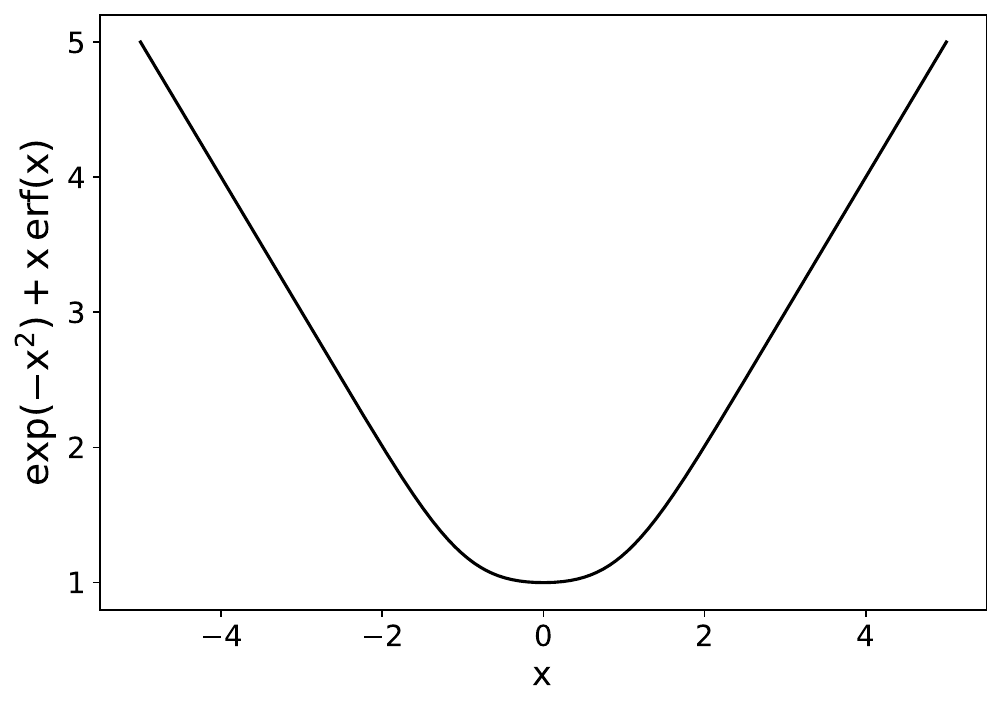}
\caption{Illustration of the functional form attained by the energy distance between two Gaussians as a function of the normalized difference between their means, $x$. For simplicity, the diagram shows the value $x$-dependent term in Eq.~\ref{dist_eq}, illustrating that it is bounded from below by $1$ and asymptotically approaches  $|x|$. The additional term of the energy-distance-based metric imposes a constant offset to the curve presented in this figure.}
\label{fig:x_dist}
\end{figure}

It is illuminating to closely examine the behavior of the expression in Eq.~\ref{dist_eq}. First, it can be shown to be always non-negative, as is required for a metric. The part that depends on $x$ is an increasing function of the absolute difference of the two means (as Fig.~\ref{fig:x_dist} demonstrates): it is larger or equal to $1$ and asymptotically converges to $|x|$. The second term, $y$, reflects the heteroscedasticity of the measurement pair. It is bounded between $2^{-1/2}$, for extreme variance ratios, to $1$, in the case of equal variances. In essence, the heteroscedasticity makes the energy-distance-based metric treat equal-variance distributions as closer to one another. 
Finally, the term in square brackets is scaled by $\sigma_{ij}$, which increases the distance between distributions with larger variances. The constant multiplicative factor $\sqrt{8/\pi}$ can be omitted in practical contexts.

Following \citet[][corollary 3.18, therein]{Lyons2013}, we take the square root of the metric to guarantee it is of `strong negative type'. Exploring the definition and intricacies of strong negative-type metric spaces is outside the scope of this paper, but it is important to note this attribute is a sufficient condition for a metric to be applicable for distance correlation computation \citep[for further details, see][]{Szeetal2007,Lyons2013}.
\begin{equation}
\mathcal{D}(F_i, F_j) = \sqrt{\varepsilon_{ij}^2}\,.
\label{distsqrt_eq}
\end{equation}
Thus we obtained a strong-negative-type metric, which we then used to formulate an adaptation of the PDC periodogram, to use measurement uncertainties (as described in the following section). 

\subsection{Phase-distance periodogram}

Following \citet{zucker18, zucker19}, we defined a distance matrix based on the pairwise energy distance introduced above. Each entry of the distance matrix represents the distance between the corresponding measurements, namely
\begin{equation}
a_{ij} = \mathcal{D}(i, j) \ .
\end{equation}
Once a distance matrix is obtained, we apply $\mathcal{U}$-centering correction \citep{sz_par_14}, 
\begin{equation}
\label{eq: u-cent}
A_{ij} = \begin{cases}
\begin{split}
a_{ij} - \frac{1}{N-2}\sum\limits_{k=1}^{N}a_{ik} 
- \frac{1}{N-2}\sum\limits_{k=1}^{N}a_{kj} \\[8pt]
+ \frac{1}{(N-1)(N-2)}\sum\limits_{k,l=1}^{N}a_{kl} 
\end{split} 
& \text{if $i \neq j$ ,} \\ \\
0 & \text{if $i = j$ .}
\end{cases}
\end{equation}
Here, we use $N$ to represent the number of measurements. The $\mathcal{U}$-centering procedure is required to produce an unbiased estimator of the distance correlation.

Thus far, we have discussed the distances between pairs of observables. To construct a periodogram, we must construct a distance matrix representing the pairwise phase distance for each trial period, $P$. We used the metric defined by \citet{zucker18} to do so. The phase distance between two measurements is 
\begin{equation}
\begin{split}
b_{ij} &= \phi_{ij}(P-\phi_{ij}) \ ,
\end{split}
\end{equation}
where
\begin{equation*}
    \phi_{ij} = (t_i - t_j)\mod P
\end{equation*}
is the phase difference between two measurements.
The resulting distance matrix is then $\mathcal{U}$-centered, in a process identical to the one presented in Eq.~\ref{eq: u-cent}, producing the $\mathcal{U}$-centered phase distance matrix, denoted as $B_{ij}$.

The unbiased estimator of the distance correlation is calculated using 
\begin{equation}
\label{eq:cor}
D = \frac{\sum\limits_{ij}A_{ij}B_{ij}}{\sqrt{\sum\limits_{ij}A^2_{ij}\sum\limits_{ij}B^2_{ij}}} \ .
\end{equation}

Similar to the original version of the PDC periodogram, the proposed recipe is computationally intensive and entails an $O(N^2)$ calculation for each trial period. This quadratic dependence may be reduced in future applications to $\mathcal{O}(N\log N)$ by using newly developed fast techniques to compute the distance correlation \citep[e.g.][]{huo2014fast,Chaudhuri_2019}.

Following \citet{Binnenfeld2023HOLDER}, we used a $\chi^2$ test to assess the significance of periods detected by the PDC \citep{fap_shen}, producing its false alarm probability (FAP). The $\chi^2$-based FAP is simple to compute and spares the need for computationally heavy permutation tests, which we previously used for this purpose \citep{binnenfeld20, Binnenfeld2022}. The FAP is calculated as follows:
\begin{equation}
\label{eq:fap}
FAP = 1 - F_{\chi_1^2 - 1}(N\cdot D)
\end{equation}
with $F$ being the $\chi^2$ distribution cumulative distribution function (CDF), $N$ is the number of measurements, and $D$ defined in Eq.~\ref{eq:cor}.

\begin{figure} 
\centering
\includegraphics[width=1\columnwidth,clip=true]{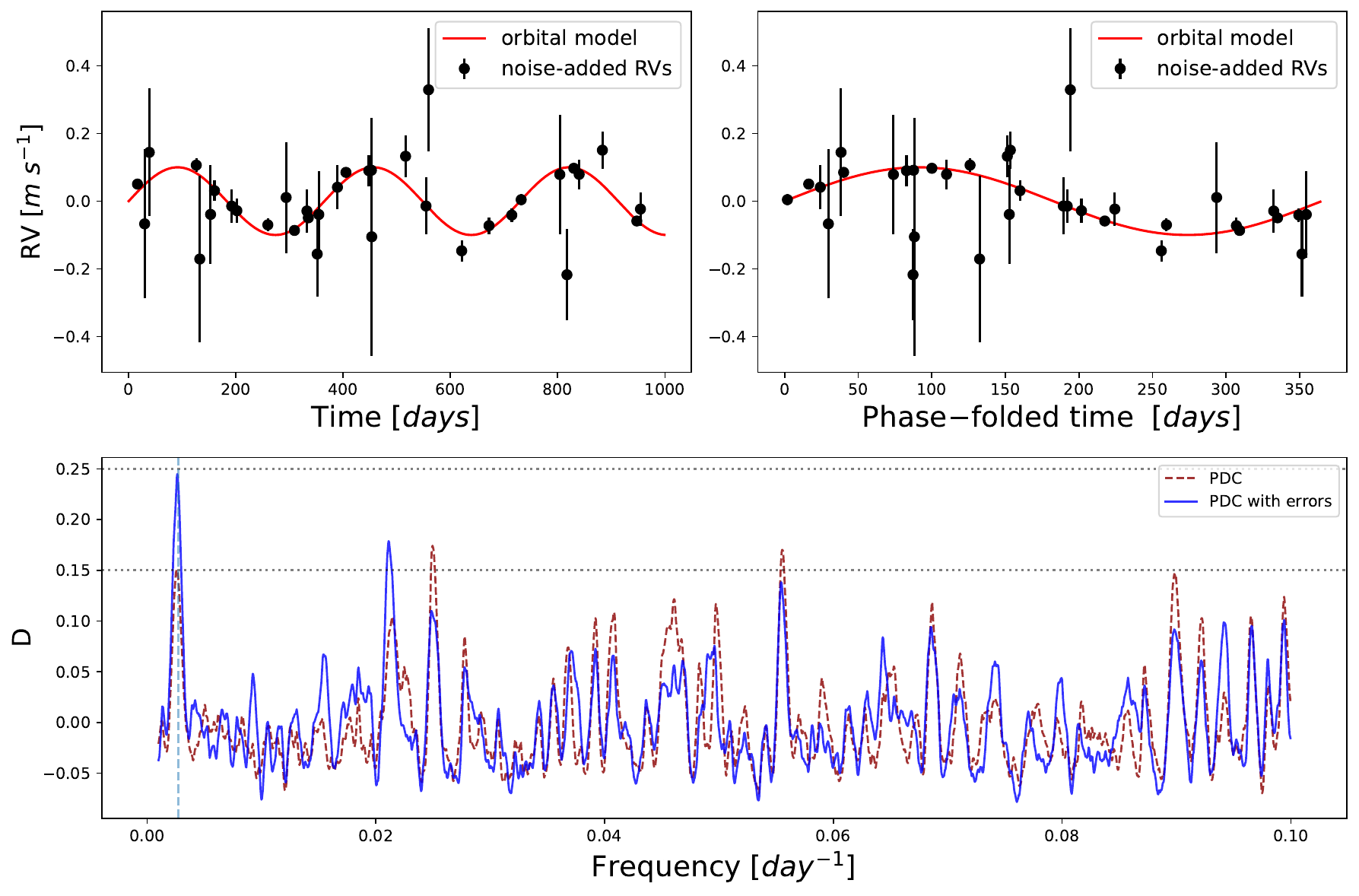}
\caption{Simulated demonstration of an Earth-analog exoplanet detection. \textit{Top:} Simulated RV curve of a planet-hosting star, as described in Sect.~\ref{sec:Lowmass}. The red line represents the orbital model, and the black circles represent the simulated noisy measurements, together with the simulated error bars. \textit{Bottom:} Comparison of the two versions of the PDC periodogram: with (solid blue) and without (dashed red) considering the error bars. The dashed vertical line marks the frequency associated with the injected orbital period of $365$ days. The lower dotted horizontal line corresponds to a FAP level of $10^{-2}$, and the upper one to a level of $10^{-3}$.}
\label{fig:earth_analog}
\end{figure}

\section{Demonstration and Verification}
\label{sec:ex2}

After introducing the theoretical foundations of the method, we demonstrate its performance using simulated and real datasets. To quantify its merit, we produced receiver operating characteristic (ROC) curves \citep[e.g.][]{FAWCETT2006861}. The ROC is a diagnostic tool for assessing the performance of detection schemes by comparing the rates of true- and false-positive detection as a function of the threshold applied to the detection statistic.

\subsection{Earth-analog exoplanet detection}
\label{sec:Lowmass}
As a first example, intended to demonstrate the capabilities and advantages of the new method for instances involving non-homogeneous errors, we generated a sinusoidal RV curve corresponding to some hypothetical exoplanet with a circular orbit of $365$ days, and radial velocity semi-amplitude of $10\,\mathrm{cm\,s}^{-1}$. The radial velocity curve was sampled in randomly drawn $35$ epochs from a uniform distribution on an interval of $1000$ days. We added noise to each measurement by randomly drawing from a normal distribution centered around $0$, with standard deviations that were drawn from an exponential distribution with a scale parameter of $1\,\mathrm{m\,s}^{-1}$. The standard deviations were paired with the corresponding simulated measurements as their `estimated' error bars. We used this procedure to produce a wide range of noise standard deviations in the same dataset. The simulated RV curve, with the simulated error bars, can be seen in Fig.~\ref{fig:earth_analog} (top panel).

As can be seen in the bottom panel of Fig.~\ref{fig:earth_analog}, the peak corresponding to the actual periodicity in the original PDC periodogram is not prominent enough compared to other spurious peaks, while the improved version of PDC does show a dominant peak at the expected annual frequency, with a FAP of $10^{-3}$.

\subsection{Detecting eccentric orbits}
Periodic RV signals related to eccentric orbits were shown to be more challenging to detect \citep{Pinamonti17}. Their diverse variability patterns pose difficulties for model-dependent methods, and they were therefore identified as preferable targets for the PDC periodogram \citep{zucker18}. 

To demonstrate how the new adaptation of PDC improves the PDC detection capabilities even further, we chose the most eccentric exoplanet currently known: HD\,20782\,b, with $e = 0.97 \pm 0.01$ and an orbital period of $597$~days \citep{otoole2008, Kane_2016}. We used RVs from the recently published HARPS radial velocity database \citep{Trifonov2020}, which consists of fifteen years of HARPS RVs.

\begin{figure} 
\centering
\includegraphics[width=1\columnwidth,clip=true]{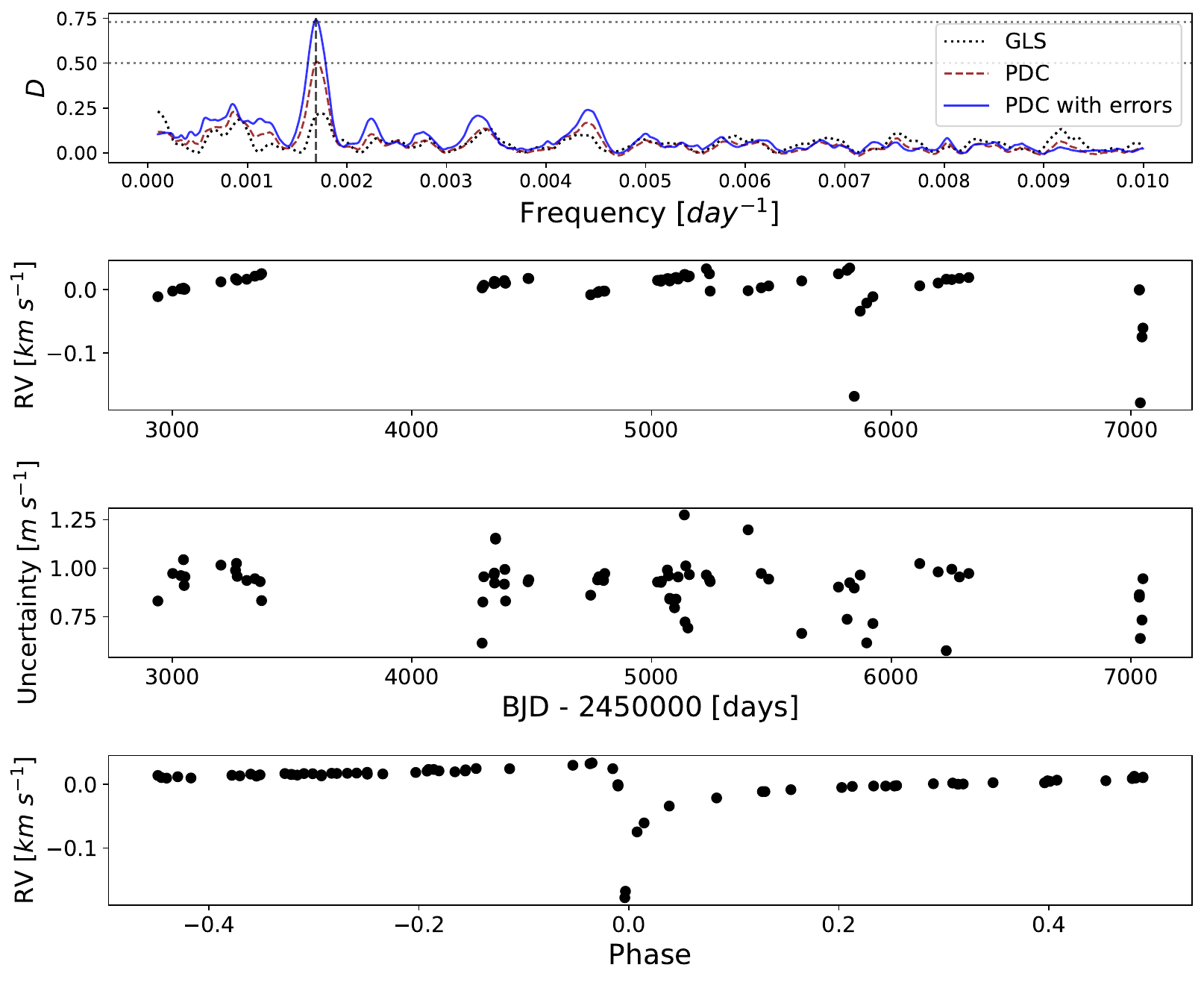}
\caption{Demonstration of the advantages of the new version of the PDC periodogram, which takes into account the uncertainties, using RV data of the planet-host star HD\,20782\,b data. \textit{Top:} GLS and PDC periodograms. The dashed vertical line marks the frequency corresponding to the known orbital period of $597$ days. The lower dotted horizontal line corresponds to a FAP level of $10^{-9}$, and the upper one to a level of $10^{-13}$. \textit{Middle panels:} HARPS RVs measurements for HD 20782, and their estimated measurement uncertainties (plotted separately for clarity). \textit{Bottom:} RVs phase-folded according to the previously published orbital period.}
\label{fig:ecc}
\end{figure}

HARPS optical fibers were upgraded in May 2015, changing the instrumental profile and introducing an RV offset between the pre- and post-upgrade RVs. \citeauthor{Trifonov2020} recommended treating the pre- and post-upgrade time series from the database as taken from two different instruments, a recommendation we indeed followed in our analysis. Out of $87$ measurements in total, we only used the $72$ taken prior to the fiber change. The RVs and their error estimates can be seen in the bottom panels of Fig.~\ref{fig:ecc}.

As seen in the top panel of Fig.~\ref{fig:ecc}, the GLS periodogram does not exhibit a real prominent peak in the previously published frequency, while the regular PDC version does. The significance of the peak is even emphasized in the improved version of the periodogram. The change translates to a major decrease in the detection FAP, from $10^{-9}$ to $10^{-13}$.

RV curves of very eccentric orbits can benefit significantly from the new adaptation of the periodogram. As demonstrated by the case of HD\,20782, samples that occur during periastron passage can easily be mistaken for outliers, even if their uncertainty estimates are small. In other cases, the error bars might be more ambiguous. The new and improved version offers an objective and quantitative way to include the error bars in the analysis

\subsection{ROC curves}

We ran the tests for the ROC curve using the same procedure described in Sect.~\ref{sec:Lowmass}, only on $1000$ sinusoidal RV curve realizations with periods drawn from a uniform distribution between $7$ and $180$ days. Similarly, we drew a random number of samples between $20$ and $70$. We added noise to each measurement by randomly drawing from a normal distribution centered around $0$, with standard deviations that were drawn from an exponential distribution with a scale parameter of $1\,\mathrm{m\,s}^{-1}$. Half of the light curves were then intentionally shuffled to eliminate the periodic signal, so that we could later estimate the FP rate. We calculated the two periodograms for each realization on the same frequency grid with 200 frequencies between $10^{-4}$ and $0.2\,\textrm{day}^{-1}$.

\begin{figure} 
\centering
\includegraphics[width=0.95\columnwidth,clip=true]{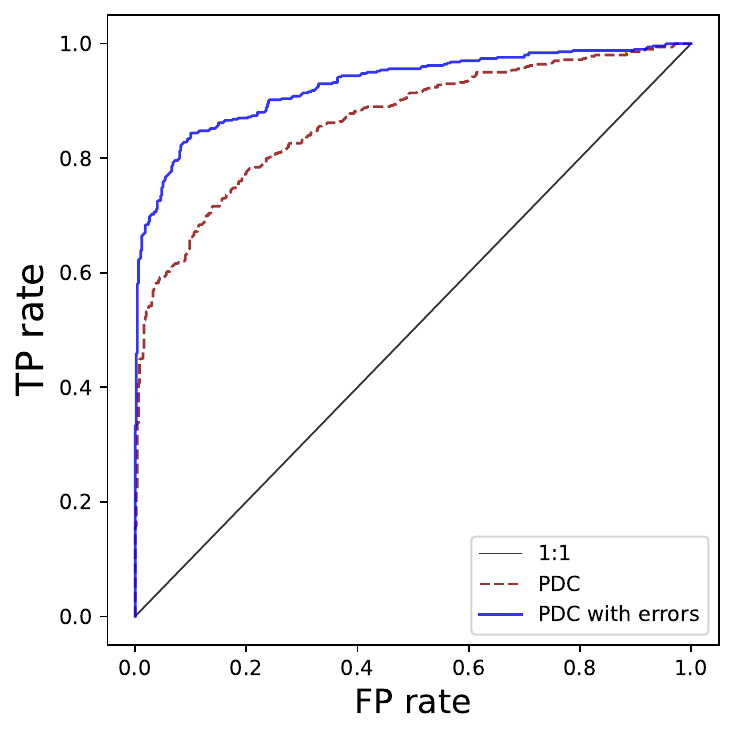}
\caption{ROC curves comparing the regular PDC periodogram with its newly developed version. The dashed red line represents the regular PDC, and the thick blue one represents the new version that considers uncertainty estimates. The thin black line stands for a $1$:$1$ ratio between the true and false positive rates, i.e., \ random guess.}
\label{fig:roc}
\end{figure}

We then used the resulting periodograms to generate the ROC curve by sorting the FAP thresholds of each periodogram in ascending order. By serially computing the pairs of FP and TP rates, we created the ROC curves presented in Fig.~\ref{fig:roc}.

As can be seen in the figure, the new version of the PDC (solid blue line), as it takes into account the uncertainties, outperforms the old version (dashed red line).

\begin{figure} 
\centering
\includegraphics[width=0.95\columnwidth,clip=true]{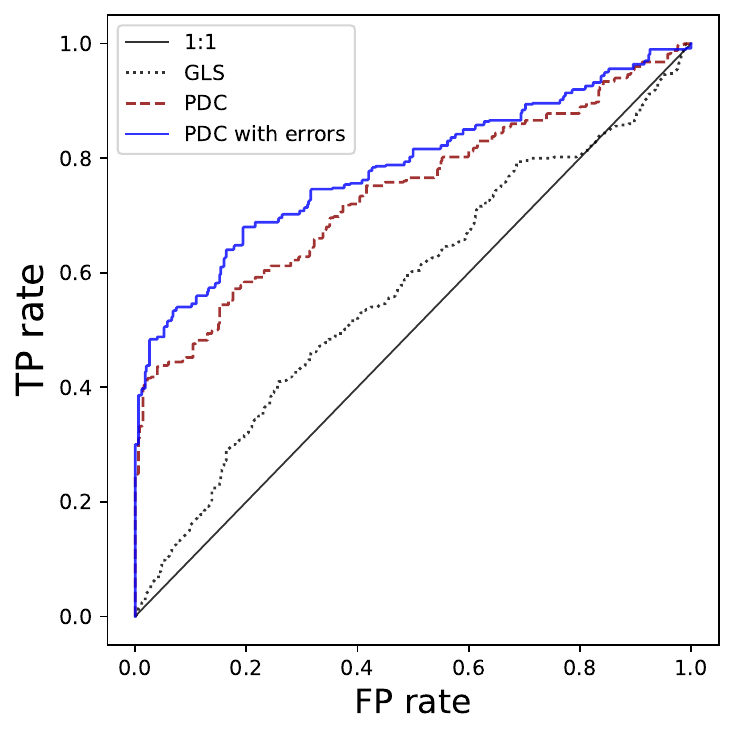}
\caption{ROC curves for the case of eccentric Keplerian RV curves ($e > 0.35$). The dashed red line represents the regular PDC, and the solid blue line represents the new version presented in this paper. They both outperform the GLS periodogram, represented by the dotted grey line. The thin black line stands for a $1$:$1$ random guess.}
\label{fig:roc_ecc}
\end{figure}

As noted in the previous subsection, the PDC periodogram outperformed other methods in detecting signals of eccentric orbits. To demonstrate this, we repeated the ROC curve test, this time simulating $1000$ realizations of eccentric Keplerian orbits (uniformly distributed $e > 0.35$). As shown in Fig. \ref{fig:roc_ecc}, the PDC periodogram in its new version (using the information contained in the error bars) does indeed outperform both the GLS and the regular version of the PDC in this detection test.

The area under the ROC curve (AUC) metric is often used for analyzing ROC curves, offering an aggregated performance measure, invariant to both scale and detection threshold. In the context of the PDC periodogram, it can be effectively employed to measure the method contribution, particularly when dealing with data exhibiting diverse characteristics such as variance and signal-to-noise ratio (S/N). Future studies, focusing on defining and exploring the physical parameter space for different test cases, could leverage the AUC metric to gain a more nuanced understanding of the periodogram efficacy.

\subsubsection{Contribution of the uncertainties information}
\begin{figure} 
\centering
\includegraphics[width=0.95\columnwidth,clip=true]{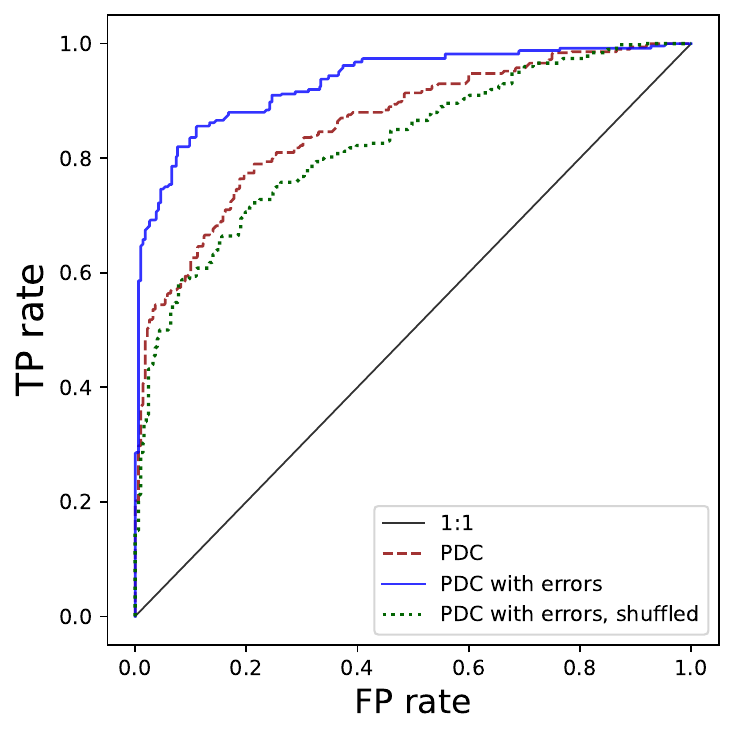}
\caption{ROC curves comparing our newly developed method in its standard configuration against a scenario where the uncertainty estimates were randomly shuffled. The dashed red line represents the regular PDC, and the solid blue line represents the new version. The green dotted line represents the performance with the randomly shuffled uncertainty estimates. The thin black line represents a random guess.}
\label{fig:roc_shffl}
\end{figure}

To further illustrate the capabilities of the updated version of the PDC and establish the importance of the contribution of measurement uncertainties to this detection scheme, we created an additional ROC curve. We repeated the $1000$ sinusoidal realization test described in the previous subsections. This time, we evaluated the performance of the periodogram under two distinct settings: the standard configuration and a scenario in which we randomly shuffled the uncertainti estimated, highlighting their critical role in the enhanced performance (presented in Fig.~\ref{fig:roc_shffl}).

As is evident from the figure, the shuffled version is outperformed by the regular PDC, while the newly developed method in its standard configuration is better than both. Nonetheless, the fact that false error estimates do not significantly impair the performance of the PDC periodogram is encouraging and attests to the robustness of the PDC.

\section{Conclusion}
\label{sec:conc}

In this paper, we present a modification of the PDC periodogram that is aimed at accounting for measurement uncertainties. We have demonstrated its performance using both simulations and data from the recently published HARPS-RVBank archive. 

The original version of the PDC periodogram ignored error bars, which was a real disadvantage of the PDC and is likely the reason the method was not widely adopted. Now that we have found a way to incorporate into PDC the information contained in the error bars, we believe we have made it much more useful.

Although this study emphasizes RV curve illustrations, it is important to note that the PDC periodogram and its current modification presented in this paper are well suited for photometric light curves as well. As we have shown \citep{zucker18}, the PDC periodogram outperforms other methods in various cases of non-sinusoidal variability patterns, such as the sawtooth-like variability shapes that are typical of pulsating stars. 

The adaptation presented in this paper relies on a Gaussian model for the uncertainty, which we used in order to construct the energy distance metric. The Gaussian uncertainty model is used very often in many contexts (almost by default) due to its simplifying properties. However, somewhat less often, other uncertainty models are used (e.g., Poisson distribution). The energy distance expression shown in Eqs.~\ref{dist_eq}-\ref{distsqrt_eq} should be replaced in those cases with the relevant expressions suitable for the assumed noise distribution. This approach may be useful in a range of applications, including the potential to extend the definition of distance to encompass and mitigate correlated noise.

In recent papers \citep{binnenfeld20, Binnenfeld2023HOLDER}, we introduced USuRPER, which is essentially a PDC periodogram for spectral observations and also a version of the PDC periodogram suitable for scanning astrometry. Despite it not being suited to the exact modification presented in this paper (i.e. scalar observations), we believe similar approaches of viewing measurements and their errors as probability distributions can be applied for those periodograms as well, to make them account for measurement error bars.

Additional relevant PDC extension is the `partial phase distance correlation periodograms', which allow for `nuisance' parameters to be accounted for with the aim of eliminating spurious peaks related to those nuisance parameters \citep{Binnenfeld2022}. In the case of two simultaneously measured scalar quantities, such as RVs and some activity indicator, the latter may be considered a nuisance; thus, it can be used to eliminate periodogram peaks originating from stellar activity. The formulation presented in this work can also be used to improve the partial periodogram by using the relevant uncertainty estimates. 
 
Finally, we provide our Python implementation of the periodogram in the form of a public GitHub repository\footnote{PDC and its extensions, including USuRPER, partial distance correlation periodograms, PDC periodogram for scanning astrometry, and the periodogram presented in this work, are all available as part of the SPARTA package \citep{SPARTA2020}, at \url{https://github.com/SPARTA-dev}.}. 

\begin{acknowledgements}
We thank the editor and the referee, Andrej Prša for reviewing this paper and for their enlightening and useful comments. This research was supported by the Ministry of Innovation, Science \& Technology, Israel (grant 3-18143) and by the ISRAEL SCIENCE FOUNDATION (grant No.~1404/22). The research of SS is supported by a Benoziyo prize postdoctoral fellowship. 

The analyses done for this paper made use of the code packages: NumPy \citep{numpy}, SciPy \citep{2020SciPy} and SPARTA \citep{SPARTA2020}.

\end{acknowledgements}

\bibliographystyle{aa}
\bibliography{pPDC}

\end{document}